\documentstyle[aps,preprint]{revtex}
\draft

\begin{document}

\title{Renormalization Group Derivation of Phase Equations}

\author{Shin-ichi Sasa}

\address{
Department of Pure and Applied Sciences, 
          University of Tokyo, 
         Komaba, Meguro-ku, Tokyo 153, Japan
\footnote{Present and permanent address}}

\address{Department of Physics, University of Illinois at Urbana-Champaign,
1110 West Green Street, Urbana, IL, 61801}

\date{Aug. 20, 1996}
\maketitle

\begin{abstract}
Phase equations describing the evolution of  large scale modulation of
spatially periodic patterns in two dimensional systems are derived by employing
the renormalization group method. A general formula for phase diffusion
coefficients is given under certain conditions.  
\end{abstract}


%% begin main text %%

%%%%%%%%%%%%%%%%%%%%%%%%% introduction %%%%%%%%%%%%%%%%%%%%%%%%%%%%
\section{Introduction}\label{sec:intro}
%%%%%%%%%%%%%%%%%%%%%%%%% introduction %%%%%%%%%%%%%%%%%%%%%%%%%%%%
% system reduction

Extracting a simpler representation of a dynamical system by restricting 
what we wish to describe, which is called system reduction, makes us 
possible to understand an apparently complicated dynamical phenomenon 
in a simple way.  In particular, perturbative system reduction is formulated 
when we can express the system in question by introducing a small parameter. 
The idea of system reduction
may have a long history since Boltzmann discussed a 
transport equation for  systems consisting of many gas molecules.
In the last three decades, systematic methods for perturbative
system reduction have been developed and applied to many examples
of dynamical phenomena appearing especially in fluid systems and reaction 
diffusion systems \cite{Hohenberg,Manneville,Kuramoto2}. 

% claim on multiple-scales

However, those who are not familiar with the methods often claim that
they seem to include a sort of art. Most artistic procedure may be 
the introduction of  multiple-scales variables \cite{Holmos}. 
Actually, the choice of a set of scaled variables in the multiple-scales 
expansion requires a guess of the final result and is justified post hoc. 
Thus, one may wish to have a standard method for perturbative system reduction 
without any post hoc justification. Such a method should be  completely 
mechanical in a sense that our intuition is excluded out. 

% befog multiple-scales

Here, we should remark that the multiple-scales technique was developed
rather recently in a history of perturbation theory on differential equations.
It is difficult to specify the origin of the perturbation theory, but 
Poincar{\'e} was the first to discuss this subject systematically 
\cite{Poincare}. The problem  was  to avoid secular terms
when a perturbative expansion is considered. By secular term  is meant 
a term unbounded with time. Subsequently, Krynov and Bogolyubov 
proposed a practically convenient and rigorously justified mathematical method
to study nonlinear oscillations \cite{Bogolyubov-M}.  The essence of 
these early important works is in the notion of a normal form of evolution 
equations. That is, by making use of a change of variables,  
evolution equations can be transformed to a simpler form. 
This notion has no relation to multiple-scales. 
Thus, one may develop a mechanical formulation relying on 
the idea of a normal form. In fact, such  perturbation theory has been 
discussed by Bogaevsky and Povzner \cite{Bogaevsky}. They developed a 
formalism generalizing the Poincar{\'e}-Bogolyubov-Krynov-Mitropolsky notion 
of a normal form. Kuramoto also attempted to formulate a perturbative system 
reduction with a geometrical interpretation based on the similar idea 
\cite{Kuramoto}. 
There seems to be no procedure of guessing a final form in their 
formulation.  A subtle point is that the existence of a well-defined 
evolution equation for new variables introduced by a change of variables 
must be assumed. Precisely speaking, it may be regarded as post hoc 
justification. 

% RG

Recently, a highly mechanical formulation for asymptotic perturbation 
theory including the perturbative system reduction has been proposed 
by Goldenfeld, Oono and their collaborators \cite{GO1,GO2,GO3}.
This was called the renormalization group (RG) approach. In their formulation, 
a naive perturbative expansion is a starting point in contrast to 
many theories which reorganize a  perturbation series 
so that secular terms do not appear.  The secular divergence is renormalized
and the  RG analysis is employed.  These procedures are well-known
in field theory. 
The RG approach has been applied successfully to many examples. 
In the context of the perturbative
system reduction, it is worthwhile noting that the obtained evolution 
equation  through the system reduction is just the RG equation. Actually,
recently, the Boltzmann equation \cite{Bolt} and 
an Euclid-invariant amplitude equation \cite{Graham}
have been derived as  RG equations. One may expect 
that universal equations 
in some sense will be regarded as renormalization group equations. 

% phase equation

This paper will give a new example supporting  this conjecture.
We will derive, by employing the RG method, 
nonlinear phase equations describing the evolution of large scale 
modulation of spatially periodic patterns in two dimensional systems.  
A phase equation for spatially periodic patterns was first derived by 
Pomeau and Manneville \cite{Pomeau}. 
Subsequently, phase equations for patterns have been derived 
in various contexts and extended to nonlinear ones \cite{Hohenberg}. 
Practically, a method developed by Cross and Newell seems most efficient
to derive phase equations \cite{Cross-Newell}. 
Cross and Newell also proposed a canonical form
of nonlinear phase equations. The phase equation we will derive is
equivalent to a Cross-Newell equation.

% outline of the paper

The outline of this paper is as follows.
In section \ref{sec:model}, we introduce  model equations to be analyzed.  
We analyze rather general partial differential equations including 
a Swift-Hohenberg (SH) equation \cite{Swift}.
In section \ref{sec:pert},
a naive perturbation will be performed. Complicated calculation of secular
terms will be shown in the Appendix A. In section \ref{sec:renorm},
we  will renormalize the 
secular terms and obtain a perturbation series uniformly valid. In section
\ref{sec:rg}, nonlinear phase equations will be obtained by the 
renormalization group 
analysis, and a general formula for phase diffusion coefficients will be
given.  In Appendix B, phase diffusion coefficients for the SH eq. 
will be calculated.  Section \ref{sec:discuss} will be devoted to discussion.

%%%%%%%%%%%%%%%%%%%%%%%%%% model %%%%%%%%%%%%%%%%%%%%%%%%%%%
\section{model} \label{sec:model}
%%%%%%%%%%%%%%%%%%%%%%%%%% model %%%%%%%%%%%%%%%%%%%%%%%%%%%

% general models

We consider partial differential equations (PDEs) of the form 
\begin{equation}
{\partial w \over \partial t}= 
F(
\{ 
 {\partial^\lambda \over \partial x_1^\lambda}
 {\partial^\mu \over \partial x_2^\mu} w
\}_{(0 \le \lambda+\mu \le M )}
) ,
\label{general}
\end{equation} 
which describes the evolution to  spatially periodic patterns 
in a two dimensional space. $w$ denotes a scalar field, and the 
generalization to the multi-component cases is straightforward.
$F$ is a polynomial of partial derivatives of $w$ up to the $M$-th
order. This implies that we do not consider an equation with a nonlocal
coupling. The assumption for $F$ will be stated in order. 
First, $F$ is assumed to be symmetric with respect to  a parity
transformation $\vec x \rightarrow -\vec x$, while the rotational invariance 
is not  assumed.  Second, as known in many cases, it is assumed that
the spatially periodic solutions form a family expressed by 
\begin{eqnarray}
w(\vec x)&=&f(\theta,\vec k), \label{solution} \\
\theta&=&\vec k \vec x +\phi,
\label{phase:theta}
\end{eqnarray}
where $f$ is a $2\pi$ periodic function in $\theta$, 
$\vec k$ is an arbitrary constant vector in a certain range
and $\phi$ is an arbitrary constant phase.
Further, for simplicity, we assume that $f$ can be chosen as 
a parity symmetric one, that is,  
\begin{equation}
f(\theta, \vec k)=f(-\theta, \vec k).
\label{symmetric}
\end{equation}
We can easily generalize our argument to the case that 
this assumption does not hold, but the analysis becomes complicated.

% set up the perturbation problem

We pay attention to a late stage in the pattern formation process, and
we assume  that  almost solutions to the equation approach to 
a neighborhood of a family of solutions expressed by Eq.(\ref{solution})
as time goes on. Then, since the solutions satisfy $F=0$, 
$\partial w /\partial t$ may be regarded as a perturbation. 
Physically, this expectation is reasonable because 
the evolution of patterns becomes slower on a later stage in the pattern 
formation process. In order to formulate a perturbation problem, 
we rewrite  Eq.(\ref{general}) as 
\begin{equation}
\epsilon {\partial w \over \partial t}= 
F(
\{ 
 {\partial^\lambda \over \partial x_1^\lambda}
 {\partial^\mu \over \partial x_2^\mu} w
\}_{(0 \le \lambda+\mu \le M )}
).
\label{epgeneral}
\end{equation} 
Do not confuse this procedure with a multiple scale
expansion method.  We have just set up the system which will be analyzed,
by restricting what we wish to describe. 
However, someone may claim that the above procedure 
is not mechanical and too formal because $\epsilon$ is not an observable 
parameter. We agree the former claim and will discuss it in the
final section. Also, the latter claim may be true in the sense that we cannot 
say the value of $\epsilon$. We consider this equation under the asymptotic
condition $\epsilon \rightarrow 0_+$. 

% example: SH equation

The simplest example of Eq.(\ref{epgeneral})
is a Swift-Hohenberg (SH) equation: 
\begin{equation}
\epsilon \partial_t w = Rw-w^3+(1-(1+\triangle)^2)w,
\label{SH}
\end{equation} 
where $R$ is a control parameter. The spatially periodic solutions 
of the SH equation are expressed by 
Eq.(\ref{solution}) with Eq.(\ref{phase:theta}) and satisfy the condition 
Eq.(\ref{symmetric}). (See Appendix B.)
Note that $M=4$ for the SH eq. 
We will develop an argument applicable to Eq.(\ref{epgeneral}).
One can consider the SH eq. as a concrete model.

%%%%%%%%%%%%%%%%%%%%%%%% naive perturbation %%%%%%%%%%%%%%%%%%%%%%%%
\section{naive perturbation} \label{sec:pert}
%%%%%%%%%%%%%%%%%%%%%%%% naive perturbation %%%%%%%%%%%%%%%%%%%%%%%%

We employ a naive perturbative expansion in $\epsilon$. 
The solution is expanded as 
\begin{equation}
w=w_0(x,t)+\epsilon w_1(x,t)+\cdots.
\end{equation}
The 0-th order solution is given by
\begin{equation}
w_0(\vec x,t)=f(\theta,\vec k(t))=f(\vec k(t) \vec x+\phi(t),\vec k(t)),
\end{equation}
where $\phi(t)$ and $\vec k(t)$ are arbitrary functions in $t$. 
Proceeding the first order in $\epsilon$, we obtain 
\begin{equation}
{\partial \phi\over \partial t}{ \partial f \over \partial \theta}
+{\partial \vec k \over \partial t}
 (\vec x {\partial f \over \partial \theta}
 +{\partial f \over \partial \vec k})
=\hat L^{(0)} w_1. 
\label{first}
\end{equation} 
Here,  $\hat L^{(0)}$ is a linearized operator around the solution
and formally expressed by 
\begin{equation}
\hat L^{(0)}= {\delta F \over \delta w}\biggr\vert_{w=f},
\end{equation}
where $ \delta / \delta w$ denotes the functional derivative along $w$.
$\hat L^{(0)}$ is  symmetric with respect to the transformation 
$\vec x \rightarrow -\vec x$ and includes partial derivative operators
up to the $M$-th order. 

Now, we solve Eq.(\ref{first}). Mathematically speaking, 
we should specify an appropriate functional space for the solutions, but 
we do not enter the mathematical details. Nevertheless, 
noting  the $\vec x$ dependence in the left-handed side of Eq.(\ref{first}),
we can expect naively that the general solutions take a form
\begin{equation}
w_1=\sum_{n=0}^{M+1} x_{(1:n)}a^{(n)}_{(1:n)},
\end{equation}
where we have introduced an abbreviation 
\begin{eqnarray}
x_{(1:n)}&=&x_{i_1}x_{i_2}\cdots x_{i_n}, \\
a^{(n)}_{(1:n)}&=&a_{i_1i_2\cdots i_n}^{(n)}.
\end{eqnarray}
We will use a similar abbreviation for  tensors with indices $i_k\cdots i_m$.
(Also, $a_{(1:0)}^{(0)}$ means a scalar variable $a^{(0)}$ and $x_{(1:0)}=1$.)
The tensor $a_{(1:n)}^{(n)}$ is a function 
in $\theta(=\vec k(t) \vec x +\phi(t))$ and in $t$ (through $t$ dependence of
$\partial \phi/\partial t$ and $\partial \vec k /\partial t$)
and can be assumed to be  symmetric  with respect to the permutation of
indices. 

Here, in order to evaluate $\hat L^{(0)} w_1$,
we define the $n$-th order tensor operator $\hat L^{(n)}$ recursively by the
formula
\begin{equation}
\hat L^{(n-1)}_{(1:n-1)}[x_{i_n}g(\vec x)]
=x_{i_n}\hat L^{(n-1)}_{(1:n-1)}g(\vec x)
  +\hat L^{(n)}_{(1:n)}g(\vec x),
\end{equation}
where $g(\vec x)$ is an arbitrary function. 
As a concrete example, 
$\hat L^{(0)}$, $\hat L^{(1)}$ and $\hat L^{(2)}$ for the SH eq.
are given by 
\begin{eqnarray}
\hat L^{(0)}&=&R-3 w_0^2-(1+\triangle)^2, \label{sh:l0}\\
\hat L^{(1)}_j &=&-4 (1+\triangle)\partial_j , \label{sh:l1}\\
\hat L^{(2)}_{ij} &=& -4(1+\triangle)\delta_{ij}-8\partial_i\partial_j.
\label{sh:l2}
\end{eqnarray}
Then, by using the operators $\{ \hat L^{(n)} \}_{n=0}^{M+1}$,
$\hat L^{(0)} w_1$ is written as
\begin{equation}
\hat L^{(0)} w_1=
\sum_{k=0}^{M+1} x_{(1:k)} \sum_{n=k}^{M+1} 
{}_nC_k  \hat L^{(n-k)}_{(k+1:n)}a^{(n)}_{(1:n)}.
\label{first:lhs}
\end{equation}
Substituting this equation to Eq.(\ref{first}), we obtain 
\begin{equation}
\sum_{k=0}^{M+1} x_{(1:k)} b_{(1:k)}^{(k)}=0,
\label{second}
\end{equation}
where 
\begin{equation}
b_{(1:k)}^{(k)}=
\sum_{n=k}^{M+1} {}_nC_k  \hat L^{(n-k)}_{(k+1:n)}a^{(n)}_{(1:n)},
\label{b:k}
\end{equation}
for $k \ge 2$, and 
\begin{equation}
b_{(1:1)}^{(1)}=
\sum_{n=1}^{M+1} n  \hat L^{(n-1)}_{(2:n)}a^{(n)}_{(1:n)}
-{\partial k_{i_1}\over \partial t}{\partial f \over \partial \theta}, 
\label{b:1}
\end{equation}
\begin{equation}
b_{(1:0)}^{(0)}=\sum_{n=0}^{M+1}   \hat L^{(n)}_{(1:n)}a^{(n)}_{(1:n)}
-{\partial k_{l} \over \partial t}{ \partial f \over \partial k_l}
-{\partial \phi \over \partial t}{\partial f \over  \partial \theta}. 
\label{b:0}
\end{equation}
Since $\vec x$ is arbitrary and $b^{(n)}$ is a periodic function in $\vec x$, 
Eq.(\ref{second}) leads  $M+2$ equations 
\begin{equation}
b_{(1:k)}^{(k)}=0,  \qquad (0 \le k \le {M+1}). 
\label{eqn:b}
\end{equation}
We can determine the tensors $\{a^{(n)}\}$ by solving the set of
equations and obtain 
\begin{equation}
w_1= (x_{lmn} \rho_{lmn}^{(3)} +x_{lm} \rho_{lm}^{(2)} +
x_{l} \rho_{l}^{(1)}) {\partial f \over \partial \theta}
+(3 x_{lm} \rho_{lmn}^{(3)} +2 x_{l} \rho_{ln}^{(2)})
{\partial f \over \partial k_n}+ 
a^{(0)}.
\label{secular}
\end{equation}
Here, $\rho_{lmn}^{(3)}$ and  $\rho_{lm}^{(2)}$ do not depend on $\vec x$
and satisfy
\begin{eqnarray}
6\rho_{lmn}^{(3)}D_{mn}&=&{\partial k_l \over \partial t}, 
\label{kai2} \\
2\rho_{lm}^{(2)}D_{lm}&=&{\partial \phi \over \partial t},
\label{kai3}
\end{eqnarray}
with
\begin{equation}
D_{mn}={1\over 2}(\Phi_0^\dagger, \hat L_{mn}^{(2)} \Phi_0)+
(\Phi_0^\dagger, \hat L_{m}^{(1)} {\partial f \over \partial k_n}),
\label{fnm}
\end{equation}
where $(u,v)$ denotes the inner product in the space
$V_{per}=\{u|u \in L_2[0,2\pi], \quad u(0)=u(2\pi)\} $, 
$\Phi_0={\partial f /\partial \theta}$ is a null eigenvector,
and $\Phi_0^\dagger$ is the adjoint null vector normalized to 
$(\Phi_0^\dagger, \Phi_0)=1$. $\rho_l$ is an arbitrary constant and $a^{(0)}$
is a periodic function with a certain arbitrariness. We do not need to know
the explicit form of $a^{(0)}$.
Eqs.(\ref{secular}) - (\ref{fnm}) are derived in  Appendix A. 

%%%%%%%%%%%%%%%%%%%%%% renormalization %%%%%%%%%%%%%%%%%%%%%
\section{renormalization} \label{sec:renorm}
%%%%%%%%%%%%%%%%%%%%%% renormalization %%%%%%%%%%%%%%%%%%%%%

The naive perturbation series up to the first order in
$\epsilon$  is summarized as 
\begin{equation}
w=f(\vec k(t) \vec x+\phi(t), \vec k(t))+\epsilon w_1,
\end{equation}
where $w_1$ is given by Eq.(\ref{secular}). 
This perturbation result does not make a sense for large $|\vec x|$ 
because of the singular behavior of $w_1$. 
However, this result becomes meaningful locally around an arbitrary 
position $\vec X$ when we successfully separate  the singular 
behavior and renormalize it to the arbitrariness in the zero-the order
solution. 
This operation corresponds to the renormalization in field theory. 

We now show that the singular behavior in $w_1$ is renormalizable.
First, we split $x_{(1:n)}$ to $x_{(1:n)}=x_{(1:n)}-X_{(1:n)}+X_{(1:n)}$,
where $x_{(1:n)}^R=x_{(1:n)}-X_{(1:n)}$ is not large 
when we focus on the system around the position $\vec X$. Then, 
the singular part in $w_1$, denoted by $w_1^{sing}$, becomes
\begin{equation}
w_1^{sing}=(X_{lmn} \rho_{lmn}^{(3)} +X_{lm} \rho_{lm}^{(2)} +
X_{l} \rho_{l}^{(1)}) {\partial f \over \partial \theta}
+(3 X_{lm} \rho_{lmn}^{(3)} +2 X_{l} \rho_{ln}^{(2)})
{\partial f \over \partial k_n}.
\label{sing} 
\end{equation}
We wish to renormalize these singular terms to the arbitrariness in
the 0-th order solution, $f(\vec k \vec x+\phi,\vec k)$. 
Let us recall that $\phi$ and $\vec k$ in $f$ are arbitrary functions in $t$.
We then define  $\phi^R(\vec X, t)$ and $\vec k^R(\vec X,t)$ by
\begin{eqnarray} 
\phi(t)&=&\phi^R(\vec X,t)+\epsilon \delta \phi(\vec X,t), 
\label{renorm:p}\\
\vec k(t)&=&\vec k^R(\vec X,t)+\epsilon \delta \vec k(\vec X,t).
\label{renorm:k}
\end{eqnarray} 
$\phi^R(\vec X,t)$ and $\delta \phi(\vec X,t)$ are called a renormalized phase 
and a counter term, respectively. $\vec k^R$ and $\delta \vec k$ are 
called similarly .
Substituting Eqs.(\ref{renorm:p}) and (\ref{renorm:k}) into $f$, we obtain
\begin{equation}
f(\vec k \vec x+\phi,\vec k)=f(\vec k^R \vec x +\phi^R, \vec k^{R})+
\epsilon[(\delta \vec k \vec x^R+\delta \vec k \vec X+\delta \phi)
{\partial f \over \partial \theta} 
+\delta  \vec k {\partial f \over \partial \vec k}].
\end{equation}
Comparing this with Eq.(\ref{sing}), we obtain the cancellation condition
of the singularities as
\begin{eqnarray}
\delta \theta + X_{lmn} \rho_{lmn}^{(3)} +X_{lm} \rho_{lm}^{(2)} +
X_{l} \rho_{l}^{(1)}&=&0, 
\label{cancell:1}\\
\delta  k_n +3 X_{lm} \rho_{lmn}^{(3)} +2 X_{l} \rho_{ln}^{(2)}&=&0,
\label{cancell:2}
\end{eqnarray} 
where we put
\begin{equation}
\delta \theta=\delta \vec k \vec X+\delta \phi.
\label{cancell:3}
\end{equation}
As a result, the renormalized perturbation series becomes
\begin{eqnarray}
w&=&f(\vec k^R(\vec X,t) \vec x+\phi^R(\vec X,t),\vec k^{R}(\vec X,t)) 
\nonumber  \\
 &+&\epsilon \{
[x_{lmn}^R \rho_{lmn}^{(3)} +x_{lm}^R \rho_{lm}^{(2)} +
x_{l}^R (\rho_{l}^{(1)} + \delta k_l) ] {\partial f \over \partial \theta} 
\nonumber \\
&+& (3 x_{lm}^R \rho_{lmn}^{(3)} +2 x_{l}^R \rho_{ln}^{(2)})
{\partial f \over \partial k_n} +a^{(0)} 
\}.
\label{renor:w} 
\end{eqnarray}
It seems strange that a singular term $\delta k_l$ remains in 
the renormalized perturbation series. However, since
it appears with the form $x_{l}^R \delta k_l$ and this term simply 
narrow the valid region of the result to 
$ |x_{l}^R \delta k_l| \sim O(1).$  
In this way, the renormalizability of the singular behavior has been shown,
and our perturbation result is meaningful {\it locally}
around the arbitrary position $\vec X$. 
Thus, obviously, we can put $\vec x=\vec X$ and then obtain
\begin{equation}
w(\vec X,t)=f(\vec k^R(\vec X,t) \vec X+\phi^R(\vec X,t),\vec k^R(\vec X,t))
+\epsilon a^{(0)}, 
\label{renor2:w} 
\end{equation}
where note that $\vec X$ can be replaced by a small letter $\vec x$ because 
$\vec X$ is arbitrary. This is a {\it globally} valid
expression  for $w$. 
Here, $\vec k^R(\vec X,t)$ and $\phi^R(\vec X,t)$ are given by
Eqs.(\ref{renorm:p}) and (\ref{renorm:k}) with Eqs.(\ref{cancell:1})
-(\ref{cancell:3}). Before discussing these equations, we consider 
a relation between $\vec k^R(\vec X,t)$ and  $\theta^R(\vec X,t)$ 
which is defined by
\begin{equation}
\theta^R(\vec X,t)=\vec k^R(\vec X,t) \vec X+\phi^R(\vec X, t).
\end{equation}
By using $\theta^R$, 
$\theta=\vec k \vec x+\phi$ is  written as
\begin{equation}
\theta=\vec k^R \vec x^R+\theta^R+\epsilon(\delta \vec k 
\vec x^R+\delta \theta).
\label{theta:thetar}
\end{equation}
Differentiating this equation with respect to $\vec X$, we obtain
\begin{equation}
\vec k^R={\partial \theta^R \over \partial \vec X},
\label{ktheta}
\end{equation}
where we have used the equalities $ \partial \theta/ \partial X_l=
\partial k_l / \partial X_m=0$ and 
\begin{equation}
\delta \vec k={\partial \delta \theta \over \partial \vec X}.
\end{equation}
The last equality  can be easily proved from Eqs.(\ref{cancell:1}) 
and (\ref{cancell:2}).
One may find that our $\theta^R$ corresponds to $\Theta$ in the
Cross-Newell's formulation for the phase dynamics \cite{Cross-Newell}.  
Actually, in the next section,  by the renormalization group analysis,
we will derive the equation for $\theta^R(\vec x,t)$, 
which corresponds to a Cross-Newll phase equation.

%%%%%%%%%%%%%%%%%%%% renormalization group %%%%%%%%%%%%%%%%%%
\section{renormalization group} \label{sec:rg}
%%%%%%%%%%%%%%%%%%%% renormalization group %%%%%%%%%%%%%%%%%%

Noting $\vec k(t)=\vec k^R(\vec 0,t)$, we rewrite Eq.(\ref{renorm:k}) as
\begin{equation}
\vec k^R(\vec 0, t)=\vec k^R(\vec X,t)+\epsilon \delta \vec k(\vec X,t).
\label{kflow}
\end{equation}
This equation is valid locally around $\vec X=\vec 0$ because of 
the singular behavior of $\delta \vec k$. (See Eq.(\ref{cancell:2}).)
Here, the capital letter $\vec X$ in the argument of renormalized 
quantities is convention. $\vec x$ can be used instead of it, of course.
We now wish to obtain a globally valid expression for $\vec k^R$. 
The RG analysis makes us possible to do it. We first show a formal
argument. Differentiating Eq.(\ref{kflow}) with respect to $\vec X$, we obtain
\begin{eqnarray}
{\partial k_l^R \over \partial X_m} &=&
 -\epsilon {\partial \delta k_l \over \partial X_m}, 
\label{kr:2} \\
                  &=& \epsilon (6X_n\rho_{lmn}^{(3)}+2\rho_{lm}^{(2)}). 
\label{kr:3}
\end{eqnarray}
Here, $\rho^{(3)}$ and $\rho^{(2)}$ are functions of $\vec k(t)$
in their definitions Eqs.(\ref{kai2}) and (\ref{kai3}), but 
we are allowed to replace $\vec k(t)$ to $\vec k^R(\vec X,t)$ 
because the difference gives the $O(\epsilon^2)$ contribution
in Eq.(\ref{kr:3}). 
Multiplying $D_{lm}(\vec k^R)$ to this equation and using Eqs.(\ref{kai2})
and (\ref{kai3}), we  obtain 
\begin{eqnarray}
D_{lm}(\vec k^R){\partial k_l^R \over \partial X_m} &=& 
\epsilon[{\partial k_l^R \over \partial t} X_l+
         {\partial \phi^R \over \partial t} ], \\
&=& \epsilon {\partial \theta^R \over \partial t}.
\label{kr:4}
\end{eqnarray}
By combining Eq.(\ref{ktheta}), this equation  can be written as 
\begin{equation}
\epsilon {\partial \theta^R \over \partial t}=D_{lm}(\vec k^R)
{\partial \over \partial X_l}{\partial \over \partial X_m}\theta^R.
\label{phase}
\end{equation}
We expect  that Eq.(\ref{phase}) is globally valid because 
it does not include a singularity. Equation (\ref{phase}) is equivalent
to a Cross-Newell equation \cite{Cross-Newell}.
However, one may claim that
Eq.(\ref{phase}) is valid only over the region where Eq.(\ref{kflow})
is meaningful because Eq.(\ref{phase}) has been derived from Eq.(\ref{kflow}).
Also, the replacement from $\vec k$ to $\vec k^R$ in Eq.(\ref{kr:3}) seems 
a little bit unnatural. 
We notice here that the RG analysis is not employed explicitly 
in the above argument. Next, we consider the problem from 
the RG point of view.

The argument in this paragraph follows a review article by Shirkov
\cite{Shirkov}. The RG is a symmetry structure with respect to 
the alternation of the way of giving boundary values
\cite{Shirkov,Stueckelberg,Bogolyubov-S}. 
Let us see this property in our system.  $Q$ denotes 
a set of renormalized quantities, i.e.
\begin{equation}
Q(\vec X)=(\theta^R(\vec X, *), \vec k^R(\vec X, *)),
\end{equation}
where $Q(\vec X)$ is a function in $t$.
{}From Eq.(\ref{kflow}) and a similar expression for $\theta^R$
derived from Eq.(\ref{theta:thetar}), we can write
\begin{equation}
Q(\vec 0)=Q(\vec X)+\epsilon \delta Q(\vec X).
\label{qflow:0}
\end{equation}
This expression is valid locally around $\vec X=\vec 0$. Using
the spatial homogeneity of the system, we can  prove 
\begin{equation}
Q(\vec X_0)=Q(\vec X)+\epsilon \delta Q(\vec X-\vec X_0),
\label{qflow:x0}
\end{equation}
which is valid locally around an arbitrary position $\vec X_0$. 
Then,  from the form of Eq.(\ref{qflow:x0}),
we assume a globally valid expression
\begin{equation}
Q(\vec X)=F(\vec X-\vec X_0, Q(\vec X_0))
\label{qflow:x0:gen}
\end{equation}
for an arbitrary position $\vec X_0$. 
Since $Q(\vec X)$ does not depend on $\vec X_0$,  
Eq.(\ref{qflow:x0:gen}) shows the invariance property of $Q(\vec X)$
under the alternation of the way of giving the boundary values.
Further, $F$ satisfies the transitivity expressed by
\begin{equation}
F(\vec X-\vec X_0, Q(\vec X_0))
=F(\vec X-\vec X_1, F(\vec X_1-\vec X_0, Q(\vec X_0)).
\label{fss}
\end{equation}
This leads to  the composition law of  
the transformation $R(\vec X)$  acting on $Q(\vec X_0)$, which is 
defined by 
\begin{equation}
R(\vec X) Q(\vec X_0)=F(\vec X, Q(\vec X_0)).
\label{trans:r}
\end{equation}
Noting that $R(\vec 0)=1$ and that $R(-\vec X)$ is the inverse transformation
of $R(\vec X)$, we can conclude that the transformations form a Lie group.
Then, according to the fundamental theorem of the Lie group theory 
\cite{Pontrjagin},
a differential equation obtained by considering an infinitely 
small transformation determines the Lie group.
The differential equation is 
\begin{equation}
{\partial Q \over \partial \vec X}=\vec \beta(Q),
\label{rge}
\end{equation}
where  $\vec \beta$ is called a generator and defined by
\begin{equation}
\vec \beta(Q)= {\partial F (\vec X, Q) \over \partial \vec X}
\biggr|_{\vec X=\vec 0}.
\end{equation}
Eqs.(\ref{fss}) and (\ref{rge}) are called functional and differential 
RG equations for the variable $Q$ respectively. 
The generator $\vec \beta(Q)$ has been referred to
as a beta function in the RG literature.

As seen in its definition, the generator is determined by the local property
of the transformation. Thus, the locally valid expression Eq.(\ref{qflow:0})
is enough to give the generator perturbatically, and through the RG equation 
Eq.(\ref{rge}), which is globally valid, we can find $Q(\vec X)$ for 
all $\vec X$.  This procedure to obtain globally improved solutions 
from locally valid ones are known as the perturbative RG method. 
{}From Eqs.(\ref{cancell:1}) and (\ref{cancell:2}), we obtain
\begin{eqnarray}
{\partial \theta^R \over \partial X_l}&=
&\beta^{(\theta)}_l(\theta^R,\vec k^R)=k^R_l, \\
{\partial k_l^R \over \partial X_m}&=
&\beta^{(\vec k)}_{lm}(\theta^R,\vec k^R)={1\over 2}(D^{-1}(\vec k^R))_{lm} 
{\partial \theta^R \over \partial t}.
\end{eqnarray}
These RG equations are equivalent to Eq.(\ref{phase}).
The procedures Eqs.(\ref{kr:2})-(\ref{kr:4}) should be regarded as 
a simplified argument of the perturbative RG method based on the Lie group
theory. Recently, Kunihiro has claimed that the RG method developed
by Goldenfeld, Oono and their collaborators is not mathematical and has 
proposed an envelop method  against the RG formulation \cite{Kunihiro}.
However, there is no necessity to consider the envelop method, 
though it may be a correct interpretation of the perturbative RG method. 
We recognize that the RG method discussed here is not purely mathematical, 
but still remains at a formal level as usual in theoretical physics. 
We believe however that a careful mathematical description will be possible. 
(See \cite{Kuwamura} for a mathematical formulation for phase dynamics.)

%%%%%%%%%%%%%%%%%%%%%%%%%%%%%%%%%% discussion %%%%%%%%%%%%%
\section{discussion} \label{sec:discuss}
%%%%%%%%%%%%%%%%%%%%%%%%%%%%%%%%%% discussion %%%%%%%%%%%%%%

We summarize the result:
The solution is expressed in the form
\begin{equation}
w=f(\theta^R,\vec k^R)+\epsilon a^{(0)}, 
\label{final} 
\end{equation}
where $\theta^R$ obeys the nonlinear phase equation Eq.(\ref{phase}), 
and $\vec k^R$ is derived from Eq.(\ref{ktheta}). 
As far as we know, the general formula for phase diffusion coefficients
has never been presented explicitly. When a model equation satisfies the 
conditions we assumed, we can immediately calculate the phase diffusion
coefficients by using the general formula Eq.(\ref{fnm}). In Appendix B, we 
show the result for the SH equation. Since many variants of the SH eq. 
\cite{Hohenberg} satisfy the conditions we assumed, this general formula 
is useful. However, there are some important classes out of our consideration.
One class consists of nonlocal models which appear commonly in fluid dynamics
because of the incompressive condition \cite{Siggia}. 
The resultant phase equation 
is coupled to mean flow and this leads to a new type of instability called   
a skewed varicose instability \cite{Cross-Newell,Cross,Newell}.
It seems that there is no technical difficulty to discuss nonlocal models,
but complication will come in. The other class out of our consideration 
consists of models which have a different type of a family of solutions.
As one example, since the Kuramoto-Sivashinsky (KS) equation \cite{KS} 
possesses a Galilei symmetry \cite{Frish}, a family of spatially periodic
solutions in the one dimensional system is expressed by 
$w=f(k(x-vt)+\phi,k)+v$, where $k$, $\phi$ and $v$ are arbitrary constants. 
Correspondingly, due to the existence of additional null mode, 
secular terms take a different form  when we employ a naive perturbative
expansion. We expect that all secular divergences are renormalized to 
$k$, $\phi$ and $v$ and the RG equation gives a correct phase equation.
We remark that  an oscillatory instability appearing 
in Rayleigh-B{\'e}nard convection with a low Prandtle number 
is associated with the Galilei invariance \cite{Fauve}.

% other type of  phase equations 

A pattern breaks the spatial translational symmetry
and the phase variable is interpreted as a Goldstone mode
for the symmetry breaking. Similarly, phase equations associated with
the temporal translation symmetry have been discussed in reaction diffusion 
systems \cite{Kuramoto2}, 
where a limit cycle solution breaks the temporal translational 
symmetry. The resultant phase equation has a similar form with a Burgurs
equation.  Renormalization group derivation of the phase equations from
general reaction diffusion equations which have a limit cycle solution
is much simpler than the derivation in this paper \cite{Sasa}, because 
the naive perturbation result in such systems shows that 
there is only one secular term proportional to $t$. 
Phase equations for propagating patterns  will be derived similarly,
though  a little complication arises  
because of the mixed nature of spatial and temporal symmetry breakings. 

% comparison with other derivation
% Cross-Newll

Let us compare the RG method with other methods for perturbative system
reduction. 
The most efficient derivation of phase equations
 may be the Cross-Newell method. 
In their formulation, besides a standard multiple-scales analysis, 
a new variable $\Theta$ is introduced  with playing two roles: 
the derivative of $\Theta$ gives the modulation of wavevectors, 
while  $\Theta$ has a multiple-scales relation with  a phase coordinate 
specifying a position of the basic periodic pattern. 
That is, their formulation  consists of a tactical combination
of the multiple-scales analysis and the method of changing  variables.
We recognize that  such an acrobatic procedure greatly reduces  
a calculation time to obtain a final form especially in complicated problems.
Their method  may be optimized for the purpose of deriving nonlinear phase 
equations. Also, the formal perturbation series can be produced 
systematically up to an arbitrary order. 

The RG derivation of phase equations need more tedious calculation
than the Cross-Newell method. 
As far as we are concerned with  simple problems such as nonlinear 
oscillations, there is no significant difference between computational 
efficiencies by the RG derivation and by conventional ones. However,
as shown in this paper, it seems obvious that a more mechanical method 
requires a more calculation time when the problem becomes complicated. 
% significance of mechanical formulation
Thus, one may doubt the practical efficiency of mechanical formulation.
Mechanical formulation seems to be apparently impractical.
However, for much complicated problems which are intractable 
by manual calculation, a more mechanical method may become more useful 
because  mechanical procedures can be programmed in computers. 
Further, mechanical methods are useful to problems 
for which  we cannot find a suitable guess of the solution. 

% \epsilon problem

We should remark that our derivation is not yet a completely mechanical 
one. As discussed in section \ref{sec:model}, there is  subtle ambiguity 
where  a small parameter $\epsilon$ is put.  We cannot formalize this 
procedure mechanically. This problem is not peculiar to the present problem,
but appears in many examples, even in a quite simple example such as 
linear ordinary differential equations  \cite{GO3}.
It may be honest to say that we set up  systems so that the RG method works 
well. 
Our choice is physically reasonable, but our purpose is to eliminate
such a phrase.   
In addition, let us recall that perturbative system reduction is possible 
only when we can express the system in question by introducing 
a small parameter. 
The formulation on this part is independent of the perturbation method
and should be discussed separately. However, we do not have a general way 
how to express what we wish to describe.  Of course, in some cases, 
we can introduce suitably a small parameter with referring to the physical 
situation in consideration, but again our purpose is to eliminate 
such a phrase.
As far as we consider in the present knowledge, 
it seems hopeless to formulate this point mechanically.

% poincare-Bogolyubov method

As mentioned in section \ref{sec:intro}, there are other theories aiming at 
a mechanical formulation. Bogaevsky and Povzner analyzed many 
examples including WKB-type problems and some PDE problems \cite{Bogaevsky}. 
The correspondence with the RG method seems clear at least in 
simple problems.  Their method is interpreted as systematic renormalized 
perturbation theory, that is, they consider  renormalized 
variables  from the outset.  The calculation may be carried out 
mechanically, but we do not check it whether or not their formulation
has the same content as the RG method  even for complicated problems
such as phase equations derived in this paper.

% kuramoto

A Kuramoto's geometrical interpretation on the system reduction
comes from the notion of a normal form \cite{Kuramoto}. 
In this sense, his formulation shares a common structure
with Bogaevsky and Povzner.  However, his main claim is that 
the system reduction is independent of choice of the method of expansion 
series. One may choose any expansion method as far as it can be 
performed consistently. In general, we can obtain a formal solution 
to a set of self-consistent equations under certain additional assumptions
which are required  from the outset in his formulation.
According to his view, when the perturbative system reduction
is considered, this formal solution is evaluated with referring to physical
situations. This is one clear strategy, but seems less practical than standard
conventional methods and less  mechanical than the RG method.  

% significance of RG 

The importance of the RG method is not restricted to its mechanical nature.
As is well known, RG has been used in diverse fields of theoretical physics
in particular since Wilson formulated RG in a different form \cite{Wilson}. 
Thus, by studying a common structure to several RG approaches, 
we may find a new method for system reduction.  

% Fss and algebraic formulation

Shirkov reviewed that RG methods appeared in several contexts were employed 
based on the transitive property of physically relevant quantities 
with respect to the alternation of the way of giving the boundary values 
\cite{Shirkov}.  
He called this property  functional self-similarity (FSS).  
The notion is almost equivalent to a group structure of the old style RG
\cite{Bogolyubov-S,Stueckelberg} 
and is a generalization of the usual self-similarity related to power laws.  
In this sense, the RG analysis is interpreted as a method to construct 
a FSS symmetric solution. 
It is not easy to know the FSS symmetry beforehand.
Nevertheless, if we find it before calculation, the group analysis 
\cite{Olver} can be
employed to an enlarged system consisting of the original equation in question
and an equation describing alteration of the boundary data. 
Such a formulation was already proposed by Kovalev, Krisvenko and Pustovalov
\cite{Kovalev}.

% Wilson-Kadanoff

The RG method employed in this paper is identical to the old style RG
developed in 1950'.  Another popular formulation of RG  
is the Wilson-Kadanoff scheme, in which the RG 
transformation is devised based on a physical insight of the system 
\cite{Wilson,Nigel}.  
Owing to its constructive nature, such a scheme may be useful in 
non-perturbative system reduction. The study in this direction is 
developing \cite{Balsera}. 
Also, related to it, the constructive RG method was employed  
to discuss asymptotic behavior of PDEs mathematically \cite{Bricmont}. 
However, we do not yet 
understand a relation among these different RG approaches.

\acknowledgments

The author is grateful to  N. Goldenfeld  for critical comments on this
study and a stimulating suggestion on future works. 
He acknowledges Y. Oono for informing the author of 
Refs.\cite{Bogolyubov-S,Kovalev} and for enlightening conversations
on the renormalization group. 
He also thanks Q. Hou for fruitful conversations on pattern formation 
problems.  
This work was done during the author's stay in University of Illinois.
He acknowledges the hospitality of the University and the support by 
the National Science Foundation grant NSF-DMR-93-14938.

%%%%%%%%%%%%%%%%%%%% Appendix %%%%%%%%%%%%%%%%%%%%%%%%%%%%%%%

\section{Appendix A: calculation of secular terms} 
%\appendix{derivation of secular terms} 

In this appendix, we solve Eq.(\ref{eqn:b}) with Eqs.(\ref{b:k})-(\ref{b:0}).
We consider the case $M=4$ which holds for the SH eq.
The extension to arbitrary $M (> 4)$ is straightforward, 
and the result is unchanged.  

The equations in question are regarded as linear
functional equations for $2\pi$ periodic functions in 
$\theta=\vec k \vec x+\phi$. 
In order to solve it, we consider an Hilbert space 
$V_{per}=\{u|u \in L_2[0,2\pi], \quad u(0)=u(2\pi)\} $, where
$\partial /\partial x_j$ in $L^{(n)}_{1:n}$ is replaced 
by $k_j \partial/\partial \theta$ 
when the operator $L^{(n)}_{1:n}$ acts on $u \in V_{per}$. 
Before solving the equations, we  summarize elementary mathematical
notions about linear algebra on the space $V_{per}$.

% null eigenvector

There exists a null eigenvector 
$\Phi_0$ satisfying
\begin{equation}
\hat L^{(0)} \Phi_0=0.
\label{null:a}
\end{equation}
In fact, differentiating the equality $F|_{w=f}=0$ with respect to
$\theta$, we obtain 
\begin{equation}
\hat L^{(0)} {\partial f  \over \partial \theta} =0.
\label{null}
\end{equation}
Thus,  we can put
\begin{equation}
\Phi_0=\partial f / \partial \theta.
\end{equation}
This null eigenvector is associated to the existence of a family of 
solutions parameterized with $\phi$ 
which comes from a spatially translational invariance of the system.  

A generalized null space in $V_{per}$ 
is spanned by only $\Phi_0$ from the following two reasons.
First, the other null mode which is 
associated with a family of solutions with  
parameterized by $\vec k$ is not included in the space $V_{per}$.
Second, if there were another null mode, a family of solutions 
would be expressed in a different form. (For a related discussion,
see section {\ref{sec:rg}.) Let us discuss the first point.
Differentiating the equality $F|_{w=f}=0$ with respect to
$\vec k$, we obtain 
\begin{equation}
 \hat L^{(0)} {df \over d\vec k}
=\hat L^{(0)} (\vec x {\partial f  \over \partial \theta}
 +{\partial f \over \partial \vec k})=0.
\label{null:wide}
\end{equation}
Then, it is easily found that the null mode 
$\vec x {\partial f  / \partial \theta}+{\partial f / \partial \vec k}$
is not a periodic function in $\vec x$.
%(This fact of course does not imply that the null mode is physically 
%irrelevant. 
%The null mode plays an important role in the long time behavior of
%the system in question and  also such a mode can be detected by numerical 
%analysis of an operator \cite{Chate}.)

% inner product

The inner product in the space $V_{per}$, which is denoted by $(u,v)$,
is defined by 
\begin{equation}
(u,v)={1 \over 2 \pi} \int_0^{2 \pi} d \theta u(\theta)v(\theta).
\end{equation}
$\hat L^{(0)}$  is not an Hermite operator in general. (Note that $L^{(0)}$ 
is an Hermite operator  for the SH eq.)  Then, an adjoint null eigenvector
is defined by
\begin{equation}
\hat L^{(0)}{}^\dagger \Phi_0^\dagger=0,
\end{equation}
where  $\Phi_0^\dagger$ is assumed to be normalized as 
\begin{equation}
(\Phi_0^\dagger, \Phi_0)=1.
\end{equation}
Consider a linear equation for $u$ in the space $V_{per}$  
\begin{equation}
\hat L^{(0)} u=b,
\label{linear}
\end{equation}
where $b$ is a $2\pi$ periodic function. Since $\hat L^{(0)}$ has a
null eigenvector, there is a solution only under the solvability condition
\begin{equation}
(\Phi_0^\dagger, b)=0.
\label{solv}
\end{equation}
When this condition is satisfied, the solution $u$ to Eq.(\ref{linear}) is 
expressed by
\begin{equation}
u=\hat M b+ \lambda \Phi_0,
\end{equation}
where $\hat M$ is a pseudo-inverse operator of $\hat L^{(0)}$ and 
$\lambda$ is an arbitrary constant. ($\hat M \hat L^{(0)}=
\hat L^{(0)} \hat M=1$ holds on a restricted space which consists of $u$
satisfying $(\Phi_0^\dagger, u)=0$. )

Here, for the later convenience, we write down an identity 
\begin{equation}
-\hat M \hat L^{(1)}_j \Phi_0={\partial f \over \partial k_j},
\label{ident}
\end{equation}
which is obtained by expressing Eq.(\ref{null:wide}) in the form
\begin{equation}
\hat L^{(1)}_j \Phi_0 +\hat L^{(0)}{\partial f \over \partial k_j}=0.
\end{equation}

Now, we derive Eqs.(\ref{secular}) - (\ref{fnm}).
We first show $a^{(n)}=0$ for $n\ge 4$.
We rewrite  Eq.(\ref{eqn:b}) with Eq.(\ref{b:k}) explicitly as
\begin{eqnarray}
\hat L^{(0)}a^{(5)}_{(1:5)}&=&0, \label{a5}\\
5\hat L^{(1)}_{(5:5)}a^{(5)}_{(1:5)}+
\hat L^{(0)}a^{(4)}_{(1:4)}&=&0, \label{a4}\\
10\hat L^{(2)}_{(4:5)}a^{(5)}_{(1:5)}+
4\hat L^{(1)}_{(4:4)}a^{(4)}_{(1:4)}+
\hat L^{(0)}a^{(3)}_{(1:3)}&=&0, \label{a3}\\
10\hat L^{(3)}_{(3:5)}a^{(5)}_{(1:5)}+
6\hat L^{(2)}_{(3:4)}a^{(4)}_{(1:4)}+
3\hat L^{(1)}_{(3:3)}a^{(3)}_{(1:3)}+
\hat L^{(0)}a^{(2)}_{(1:2)}&=&0, \label{a2}
\end{eqnarray}
{}From Eq.(\ref{a5}), we obtain
\begin{equation}
a^{(5)}_{1:5}=\rho_{1:5}^{(5)} \Phi_0,
\label{a5:sol}
\end{equation}
where $\rho_{1:5}^{(5)} $ is a constant whose value will be determined later.
We substitute Eq.(\ref{a5:sol}) to Eq.(\ref{a4}). 
Then, the solvability condition for $a^{(4)}$ is satisfied due to the parity
symmetry, and we obtain $a^{(4)}$: 
\begin{equation}
a^{(4)}_{1:4}=\rho_{1:4}^{(4)} \Phi_0
+5\rho_{1:5}^{(5)}{\partial f \over \partial k_{i_5}},
\label{a4:sol}
\end{equation}
where we have used Eq.(\ref{ident}), and 
$\rho_{1:4}^{(4)} $ is a constant whose value will be determined later.
We substitute Eqs.(\ref{a5:sol}) and (\ref{a4:sol}) to Eq.(\ref{a3}).
Then, the solvability condition for $a^{(3)}$ yields
\begin{equation}
\rho^{(5)}_{1:5}D_{4:5}=0,
\end{equation}
where 
\begin{equation}
D_{mn}={1\over 2}(\Phi_0^\dagger, \hat L_{mn}^{(2)} \Phi_0)+
(\Phi_0^\dagger, \hat L_{m}^{(1)} {\partial f \over \partial k_n}).
\label{fnm:a}
\end{equation}
Using the regularity of the matrix $D$ which is assumed,
we obtain 
\begin{equation}
\rho^{(5)}_{1:5}=0.
\end{equation}
Under the solvability condition, we obtain $a^{(3)}$: 
\begin{equation}
a^{(3)}_{1:3}=\rho_{1:3}^{(3)} \Phi_0
+4\rho_{1:4}^{(4)}{\partial f \over \partial k_{i_4}},
\label{a3:sol}
\end{equation}
where we have used Eq.(\ref{ident}), and 
$\rho_{1:3}^{(3)} $ is a constant whose value will be determined later.
We substitute Eqs.(\ref{a5:sol})-(\ref{a3:sol}) to Eq.(\ref{a2}).
Repeating the same argument, we obtain
\begin{equation}
\rho^{(4)}_{1:4}=0,
\end{equation}
and
\begin{equation}
a^{(2)}_{1:2}=\rho_{1:2}^{(2)} \Phi_0
+3\rho_{1:3}^{(3)}{\partial f \over \partial k_{i_3}}.
\label{a2:sol}
\end{equation}
One can easily see that $a^{(n)}=0$ ($n \ge 4$) even for the case
$M >4$.
Then,  Eq.(\ref{eqn:b}) with Eqs.(\ref{b:1}) and (\ref{b:0}) becomes
\begin{eqnarray}
3  \hat L^{(2)}_{(2:3)}a^{(3)}_{(1:3)}+
2  \hat L^{(1)}_{(2:2)}a^{(2)}_{(1:2)}+
 \hat L^{(0)}a^{(1)}_{(1:1)}
&=&{\partial k_{i_1} \over \partial t}{\partial f \over \partial \theta},
 \label{a1} \\
\hat L^{(3)}_{(1:3)}a^{(3)}_{(1:3)}+
\hat L^{(2)}_{(1:2)}a^{(2)}_{(1:2)}+
\hat L^{(1)}_{(1:1)}a^{(1)}_{(1:1)}+
\hat L^{(0)}a^{(0)}
&=&
{\partial k_{i_1} \over \partial t}{ \partial f \over \partial k_l}
+ {\partial \phi \over \partial t}{\partial f \over \partial \theta} 
\label{a0}
\end{eqnarray}
The solvability condition for $a^{(1)}$ yields
\begin{equation}
6\rho_{1:3}^{(3)}D_{2:3}={\partial k_{i_1} \over \partial t} 
\label{kai3:a}
\end{equation}
Under the solvability condition, we obtain $a^{(1)}$: 
\begin{equation}
a^{(1)}_{1:1}=\rho_{1:1}^{(1)} \Phi_0
+2\rho_{1:2}^{(2)}{\partial f \over \partial k_{i_2}},
\label{a1:sol}
\end{equation}
where we have used the parity symmetry and Eq.(\ref{ident}). 
$\rho_{1:1}^{(1)} $ is a constant whose value is {\it not} determined.
Repeating the same argument for Eq.(\ref{a0}), we obtain
\begin{equation} 
2\rho_{1:2}^{(2)}D_{1:2}={\partial \phi \over \partial t}.
\label{kai2:a}
\end{equation}
$a^{(0)}$ is unnecessary for our argument, though we can calculate it.

\section{Appendix B:phase diffusion coefficients}
%\appendix{phase diffusion coefficients}

In this appendix, we derive concrete expressions of phase diffusion
coefficients for the SH eq. based on the general formula Eq.(\ref{fnm}).  
A spatially periodic solution of the SH eq.  is expressed by
\begin{equation}
w(x)=f(\theta,\vec k)=A_1(k) \cos\theta+A_3 (k)\cos\theta+\cdots,
\end{equation}
where $\theta=\vec k \vec x+\phi$, and
\begin{equation}
A_1={2 \over \sqrt{3}}\sqrt{R-(1-k^2)^2}.
\end{equation}
Since $ A_3 \sim O(R^{3/2})$ for $R \rightarrow 0$, we consider only
one mode. Hereafter, we will not use the smallness of $R$, but keep
in mind that the validity is restricted to the small $R$. If someone wishes
to obtain phase diffusion coefficients for  large $R$, he needs to write
a computer program. However, main features of the phase diffusion coefficients
can be obtained under this assumption. 
Then,  $\Phi_0$, $\partial f / \partial k_l$, and $\Phi_0^\dagger$ are
easily given by
\begin{eqnarray}
\Phi_0&=&-A_1\sin\theta, \\
{\partial f \over \partial k_l}&=
&{2 (1-k^2)k_l \over R-(1-k^2)^2}A_1\cos\theta, \\
\Phi_0^\dagger&=&-{1 \over \pi A_1} \sin\theta.
\end{eqnarray}
Substituting these expressions and $\{ \hat L^{(n)}\}_{n=0}^2$ 
given by Eqs.(\ref{sh:l0}) and (\ref{sh:l2}) to the general formula
Eq.(\ref{fnm}), we obtain 
\begin{equation}
D_{nm}=-2(1-k^2)\delta_{nm}+4k_nk_m-{8(1-k^2)^2k_nk_m \over R-(1-k^2)^2}.
\end{equation}
Since the system has a rotational symmetry, $D_{nm}$ is further written as
\begin{equation}
D_{nm}=D_{//}(k^2){k_nk_m \over k^2}+D_{\perp}(k^2)
(\delta_{nm}-{k_nk_m \over k^2}),
\end{equation}
where 
\begin{eqnarray}
D_{//}&=&6k^2-2-{8(1-k^2)^2k^2 \over R-(1-k^2)^2}, \\
D_{\perp}&=&-2(1-k^2).
\end{eqnarray}

%%%%%%%%%%%%%%%%%%%%%%%%%%%%%%%%%%%%%%%%%%%%%%%%%%%%%%%
%               references                            %
%%%%%%%%%%%%%%%%%%%%%%%%%%%%%%%%%%%%%%%%%%%%%%%%%%%%%%%

\end{document}